\documentclass[fleqn,usenatbib]{mnras}

\usepackage{newtxtext,newtxmath}

\usepackage[T1]{fontenc}
\usepackage[utf8]{inputenc}
\usepackage{ae,aecompl}

\usepackage{graphicx}   
\usepackage{amsmath}    
\usepackage{amssymb}    

\usepackage[usenames,dvipsnames]{color}

\newcommand{\saxj}{SAX J1808.4-3658}
 
\title[Evolution of \saxj]{Evolutionary paths of binaries with a neutron star. I. The case of \saxj}

\author[Tailo, M. et al.]{
M.Tailo$^{1,4}$\thanks{E-mail: mrctailo@gmail.com},
F. D'Antona$^{2}$,
L. Burderi$^{1,5}$,
P. Ventura$^{2}$,
T. di Salvo$^{3}$,
A. Sanna$^{1}$,
\newauthor
A. Papitto$^{2}$,
A. Riggio$^{1}$,
A. Maselli$^{1}$
\\
$^{1}$Dipartimento
di Fisica, Università degli Studi di Cagliari, SP Monserrato-Sestu km 0.7, I-09042 Monserrato, Italy\\
$^{2}$INAF – Osservatorio Astronomico di Roma, Via di Frascati 33, I-00044 Monteporzio Catone (Roma), Italy\\
$^{3}$Dipartimento di Fisica e Chimica, Università degli Studi di Palermo, via Archirafi 36, I-90123 Palermo, Italy\\
$^{4}$Dipartimento di Fisica e Astronomia ``Galileo Galilei'', Univ. di Padova, Vicolo dell'Osservatorio 3, Padova, IT-35122\\
$^{5}$INFN, Sezione di Cagliari, Cittadella Universitaria, I-09042 Monserrato, Cagliari, Italy}
\date{Accepted 2018 June 15. Received 2018 June 15; in original form 2018 April 17}

\pubyear{2017-2018}

\begin{document}
\label{firstpage}
\pagerange{\pageref{firstpage}--\pageref{lastpage}}
\maketitle

\begin{abstract}
The evolutionary status of the low mass X--ray binary \saxj\ is simulated by following the binary evolution of its possible progenitor system through mass transfer, starting at a period of $\sim$6.6\ hr. The evolution includes angular momentum losses via magnetic braking and gravitational radiation. It also takes into account the effects of illumination of the donor by both the X--ray emission and the spin down luminosity of the pulsar. The system goes through stages of mass transfer and stages during which it is detached, where only the rotationally powered pulsar irradiates the donor. We show that the pulsar irradiation is a necessary ingredient to reach \saxj\ orbital period when the donor mass is reduced to 0.04--0.06\,M$_\odot$. We also show that the models reproduce important properties of the system, including the orbital period derivative, which is shown to be directly linked to the evolution through mass transfer cycles. Moreover  we find that the effects of the irradiation on the internal structure of the donor are non negligible, causing the companion star to be non completely convective at the values of mass observed for the system and significantly altering its long term evolution, as the magnetic braking remains active along the whole evolution. 
\end{abstract}




\section{Introduction}
\label{INTRO}

\saxj\ has been the first low mass X--ray binary (LMXB) discovered as an accreting millisecond pulsar (MSP),  showing persistent 401\ Hz pulsations during outburst \citep{wijnands-vanderklis1998}, so leading support to the recycling scenario for the spin up of old, slow rotating neutron stars \citep[NS,][]{bhattacharya_1991}. First discovered by BeppoSAX in 1996 \citep{zand_1998}, the short orbital period of this system \citep[2\ hr,][]{chakrabarty_1998}, together with the plausible mass of its donor component, in brown dwarf regime \citep[$\sim$0.05$M_\odot$, e.g.][and references therein]{sanna_2017}, shows that its secular evolution must have been driven by angular momentum losses (AML) due to magnetic braking (MB) and gravitational radiation (GW) which cause orbital variation and mass transfer from the main sequence donor. Also the binary must have become semi-detached at an orbital period shorter the the so called bifurcation period \citep[e.g.][]{ergmasarna1996}. Recently, attention to evolution has been renewed by the discovery of few systems swinging between the accretion powered status (AMXP) and the rotation powered emission (radio MSP) \citep{papitto2013,papitto_2015} confirming the link between the two stages, but raising further questions about the processes at play. 

Concerning secular evolution, it is well known that the evolution of a binary with a neutron star as primary component must be very different from the standard model, well tested for the evolution of binaries in which the primary is a white dwarf \citep[cataclysmic variables, CVs, e.g.][]{hameury1988}; in fact, for a same orbital period, the donor components of LMXBs have, generally, masses much smaller than their counterparts in CVs, indicating that their radius is more bloated. 
For instance, the donor star of \saxj\ has a mass well below 0.1$M_\odot$ \citep[e.g.][ and references therein]{sanna_2017}, while a value closer to 0.2$M_\odot$\ is predicted for a CV donor just below the 3--2$hr$\ period gap. It is well understood that an important role in the evolution is played by the irradiation of the donor due to the X--ray emission \citep{pod_1991,dantona_1994} or by the pulsar radiation in MSP binaries \citep{dantona_1993}. Recent modelling of the whole secular evolution including these effects \citep[e.g][]{benvenuto_2012,benvenuto_2014,benvenuto_2017} has shown that long periods of mass transfer (where the system is observed as a LMXB) and detached periods (in which the accelerated NS may appear as MSP) are alternated. 
The time-scales of these cycles are very long ($\gtrsim 10^6\ \rm yr$) so this behaviour has nothing to do with the short phases of alternance between LMXB and MSP stages shown by transitional MSP \citep{papitto_2015}; nevertheless, it is mandatory to have a global description of secular evolution.

Fundamental parameters in this sense are the system orbital period and the mass function. The latter provides an estimate of the donor mass, once reasonable limits to the inclination are established and a mass for the primary NS is assumed. In addition, for MSP --detached-- systems, it is important to have an estimate of how much the former donor Roche lobe is filled. In order to describe the secular evolution, the mass accretion rate $\dot M_{\rm accr}$\ (during outbursts, or averaged on longer time-scale) for X--ray sources, and the orbital period derivative $\dot P_{\rm orb}$\ are precious information. When we assume that the secular evolution of the system is in a stationary stage, the mass transfer rate and orbital period derivative are a result of the system AML, of the fraction of mass lost by the donor which results lost by the system too, and of the associated loss of specific angular momentum \citep[see e.g.][]{disalvo_2008}. When the evolution takes into account the consequences of irradiation on the donor evolution, both $\dot M_{\rm accr}$\ and $\dot P_{\rm orb}$\ do not directly provide such information, as their values (and even the sign of the period derivative) depend dramatically on the time at which we are looking at the system, within the irradiation cycle. Further, other parameters came into play, such as the strength of the pulsar magnetic field and its evolution, as the pulsar power determines the irradiation during the radio MSP stages.

Being the first accreting MSP observed, and having undergone numerous outbursts in recent years, \saxj\ has a rich observational history. Indeed many authors, based on the last few outbursts observed, gave estimates about the value of some crucial physical parameters that, alongside its orbital ones, help to understand the current state and the previous evolutionary path of \saxj. 
More specifically this system shows a remarkably high orbital period derivative ($\dot P_{\rm orb}$); indeed \cite{burderi_2006,burderi_2009,disalvo_2008,sanna_2017} measured $\dot P_{\rm orb}\sim 3.6 \times 10^{-12} s/s$ which implies a similarly high mass transfer rate ($\dot M$). The same authors at this regard gave an estimate of  $\dot M \sim -2\times 10^{-9}\ M_\odot/yr$. These estimates have been obtained with the assumptions that the secular evolution of the system is guided by AML via GW, that mass transfer would be highly non-conservative and that the donor star has a mass radius relation index of --1/3.   \cite{hartman_2008,hartman_2009, patruno_2012,patruno_2017b} observed similar values for $\dot P_{\rm orb}$, but their explanation involve tidal quadrupole interaction \`a la \cite{applegate_1992} and \cite{applegate_1994}\footnote{A scenario also explored by \cite{sanna_2017}.}. As a consequence of the high mass transfer rate observed, the spin frequency derivative of the pulsar ($\dot \nu _{\rm psr}$) during the outbursts is in the range  $\dot \nu _{\rm psr,b} = 7 \times 10^{-14}  \div 5\times 10^{-13} \ Hz/s$; where the subscript \textit{b} indicates the outburst phase.

Regarding other pulsar parameters, different observers have given slightly different estimates about the magnetic field (B) of the primary component. \cite{cackett_2009} and \cite{papitto_2009} give B=$2 \div 4 \times 10^8\ G$; \cite{burderi_2006} and \cite{sanna_2017} give B$\sim 2\times 10^8$ and B= $2 \div 3 \times 10^8\ G$, respectively. Similar values have been also found by \cite{hartman_2008,hartman_2009,patruno_2012}. The values for B imply a secular value for $\dot \nu _{\rm psr}= -6.2 \times 10^{-16} \div -1.5 \times 10^{-15}$\,Hz/s, considering all the sources listed before.    
  
 In this work, we show newly calculated irradiated LMXB models in order to describe the present status of \saxj\ and its possible evolutionary path, starting from a hypothetical progenitor system from the time it became a semi-detached binary. The irradiation model we adopt is a bit different to the one used in the literature \citep[e.g][and subsequent works]{benvenuto_2012}, as it will include both the X radiation and the MSP contribution in the calculation of the irradiation luminosity, the latter usually neglected in the calculation as it is few orders of magnitude lower than the former (during the x-ray active phases). We will also show how this contribution is crucial for the peculiar secular evolution of these systems. The present work is divided into three main parts: in \S~\ref{MODELS} we will present and describe the adopted recipes; \S~\ref{EVO} will be devoted to describe the general behaviour of the models evolution. More in detail, \S~\ref{EVO_GEN} will describe the details of the evolutionary phases they go through, with a description on how selected physical parameters of the companion stars  (\S~\ref{EVO_DONOR}) and of the NS  (\S~\ref{EVO_NS}) evolve; finally in \S~\ref{EVO_1808} we will present a comparison with the observation of \saxj. \S~\ref{Disc} will host a summary of our findings and our conclusions.

\section{Models}
\label{MODELS}

To describe the evolutionary paths that links \saxj\ to its previous history we take advantage of the capabilities of the stellar evolution code ATON 2.0 \citep{ventura_1998,mazzitelli_1999,ventura_2008} in its binary version  \citep{dantona_1989,schenker_2002,king_2002,lavagetto_2004,lavagetto_2005}. 

The basis and main capabilities of the code in its isolated star version are fully described in \cite{ventura_1998} and \cite{mazzitelli_1999}, and the interested reader can find there the choices of opacities tables, equation of state and convection model available. 
We will describe here the main assumptions made to model mass transfer during binary evolution and the recipes used to simulate the irradiated LMXB evolution; with the note that those inputs not described here have been left untouched in comparison to the version used in \cite{schenker_2002,king_2002,lavagetto_2004,lavagetto_2005}, and references therein.

\subsection{Binary mass transfer and angular momentum loss}
\label{MODEL_BIN}
The evolution of a binary towards a semi--detached configuration, occurs either due to the  increase of the stellar radius ($R_2$) of the secondary component of mass $M_2$\ due to stellar evolution, or to the decrease of the Roche lobe radius $RR_2$\ due to systemic angular momentum losses.
The code computes the approach of the donor to mass loss explicitly, according to the formulation by  \cite{ritter_1988}:

\begin{equation}
\dot M \propto exp\left[ \frac{R_2-RR_2}{H_{\rm P}} \right]
\end{equation}

where $H_{\rm P}$ represents its pressure scale height.  This is useful for any binary evolution \citep{ritter_1988,dantona_1989}, but demanded when irradiation is included. We take the opportunity to comment on the possibility that the star radius can in fact exceed its Roche lobe radius\footnote{Indeed this is the case of the models shown in \S~\ref{EVO}} as we do not impose a strict contact condition; indeed in the \cite{ritter_1988} formulation the condition to obtain a semi-detached system is $R_2 \simeq RR_2$. 
A fraction $\eta_{\rm accr}$\ of the mass lost is accreted onto the primary component. In this work we adopt $\eta_{\rm accr}=0.5$, and assume that half of the mass lost from the donor is lost from the system carrying away its angular momentum \cite[][and references therein]{schenker_2002,king_2002}. We limit the accretion to the Eddingtion limit and if it exceeds it, we lower the value of  $\eta_{\rm accr}$\ to have a rate at most equal to the limit.

Orbital angular momentum loss by gravitational waves and magnetic braking are included. To model GW angular momentum loss we refer to \cite{taylor_1979}. We adopt the \cite{verbunt_1981} model for the MB, where we use a value of 1.0 for the free parameter inside the formula. 
When following the secular evolution of cataclysmic variables, it is generally assumed that the AML due to MB switches off when the donor becomes fully convective for the first time. In this work, we maintain MB active for the whole evolution. The reason for this choice will be fully explained in \S~\ref{EVO_DONOR}. 

\subsection{Irradiation luminosity and NS evolution}
\label{MODEL_IRR}

To describe the source of the energy irradiating the secondary component in the system, we adopt a two components model: the luminosity of the pulsar itself and the X-ray radiation produced when mass transfer is active. 

The luminosity of the pulsar ($L_{psr}$) is obtained through Larmor's formula. We have then:

\begin{equation}
L_{psr}=\frac{2}{3c^3}\mu^2 \Omega^4
\end{equation}

where $\mu$ is the magnetic momentum of the pulsar in $G\times cm^3$; $\Omega$ is the pulsar frequency, and $c$ is the speed of light. 
If we use the equation for the time variation of the rotational energy ($\dot E = I \Omega \dot \Omega$) we re-obtain the familiar expression for the pulsar dipolar spin down ($\dot \Omega_{\rm dip}$)

\begin{equation}
\dot \Omega_{\rm dip} = - \frac{5}{3c^3 }  \mu^2 \Omega^3 \frac{1}{MR^2}
\label{EQN_PDIP}
\end{equation}

where M and R are the mass and the radius of the pulsar, respectively. For the purpose of this work, we are keeping the NS radius at a constant value of $10 \rm km$.

The mass accretion rate $\dot M_{\rm accr}$ on the NS is responsible for the X-ray radiation emitted. The X luminosity ($L_X$) generated by the transfer is then:

\begin{equation}
L_X= \frac{GM\dot M_{\rm accr}}{R}.
\end{equation}

$\dot M_{\rm accr}$ onto the NS cedes its angular momentum to the primary ($\dot \Omega_{\rm accr}$), according to:

\begin{equation}
\frac{\dot \Omega_{\rm accr}}{\Omega} = \frac{\dot M_{\rm accr}}{M} \left[ \frac{5}{2} \frac{\omega_{k}}{\Omega}\left( \frac{R_{\rm mag}}{R} \right)^{\frac{1}{2}} -1 \right]
\end{equation}


where $R_{\rm mag}$ is the magnetosferic radius, $\omega_k$ is the keplerian angular velocity

\begin{equation}
\omega_k=\left(\frac{GM}{R^3}\right)^\frac{1}{2}.
\end{equation}

This contribution is obtained through consideration on angular momentum and energy conservation and, in principle can be negative. With these two components in the evolution of the pulsars spin we have that the final rate is $\dot \Omega = \dot \Omega_{\rm dip} + \dot \Omega_{\rm accr}$.
The way we determine $\dot M_{\rm accr}$  is the same as in \cite{lavagetto_2004} and \cite{lavagetto_2005}, where we consider accretion if $R_{\rm mag} < R_{\rm co}$; the former is the magnetospheric radius and the latter is the corotation radius. 

We do not include evaporation \citep[usually evaluated according to][]{stevens_1992} in our models, contrarily to other works dealing with long term evolution \citep[e.g.][and references therein]{tauris_1999,buning_2004,benvenuto_2012,chen_2013,benvenuto_2014,chen_2017}.
  Indeed, the contribution from the pulsar spin down luminosity is usually neglected in the calculations of the irradiation field while at the same time only used to compute the evaporation effects, since is a few orders of magnitude lower than the the one of the X-ray radiation (see later sections). By not including evaporation at this stage we have the possibility to study the direct effects of this contribution on the donor structure and the system evolution. 

Our simulations include the treatment of radio ejection according to \cite{burderi_2001}: if the period of the system will exceed the critical period predicted ($P_{crit}$), obtained following Eq. 2 in 
\cite{burderi_2001}, written below:
\begin{equation}
\begin{aligned}
P_{crit}=1.05(\alpha^{-36} n_{0.615}^{-40} R_6^{34})^{3/50}L_{36}^{51/25}M^{1/10}\mu_{26}^{-24/5}P_{-3}^{48/5}\\ \times \left[ 1-0.462\left( \frac{M_2}{M+M_2}\right)^{1/3} \right]^{-3/2} (M+M_2)^{-1/2}\ hr;  
\end{aligned}
\label{EQN_REJ}
\end{equation}  
the system ejects the whole mass lost from the donor, as well as its angular momentum. Here $\alpha$ is the Shakura-Sunayev viscosity parameter, $n_{0.615}$ is the mean particle mass in proton unit (for a solar mixture $n_{0.615} \sim 1$) and $R_6,\ L_{36}$ and $\mu_{26}$ are the radius, the luminosity and the magnetic momentum of the NS in units of $10^6\ cm, 10^{36} erg/s$ and $10^{26}\ G\ cm^3$, respectively and $P_{-3}$ is the pulsar spin period in ms.

We also describe the evolution of the magnetic field of the pulsar by adopting a simple accretion induced decay, where the magnetic momentum follows the evolution given by \citep{shiba_1989}: 

\begin{equation}
\mu=\frac{\mu_i}{1+\Delta M_{\rm accr}/m_B}
\end{equation} 

where we adopt $m_B = 0.30\ M_\odot$ for the constant inside the model.

After we compute both L$_x$ and L$_{psr}$, we can calculate the total radiation  incident onto the companion star (or heating luminosity, $L_{h}$) as:

\begin{equation}
L_{h} = \frac{R_2^2}{4\pi a^2} (\epsilon_x L_x + \epsilon_{psr} L_{psr}) =\frac{R_2^2}{4\pi a^2}\epsilon (L_x +L_{psr})
\end{equation}
where $\epsilon_x $  and $ \epsilon_{psr}$ are two free efficiency parameters we use to regulate irradiation and $a$ is the orbital separation. 
In principle, the value for both processes could be different but, in order to not have our description dependent on too many free parameters at this stage, we will assume an equal value for both (so $\epsilon_x = \epsilon_{psr} = \epsilon$) . 

After all the coefficients have been applied and the effective heating luminosity has been computed we proceed to model the evolution of the companion star, following the procedure outlined in \cite{dantona_1993}, hereinafter DE93, with  differences explained below. 
We associate to $L_h$ an heating temperature ($T_h$), defined as follows:

\begin{equation}
T_h=\left( \frac{L_h}{4 \pi \sigma R_2^2} \right)^{1/4},
\label{EQN_TH} 
\end{equation}

where $\sigma$ is the Stefan-Boltzmann constant. Also in the DE93 scheme the final luminosity ($L$) of the companion star would be

\begin{equation}
L=L_{int} - L_h = 4 \pi R_2^2 \sigma (T_{int}^4 - T_h^4),
\label{EQN_L}
\end{equation}

Where the subscript \textit{int} refers to the intrinsic values for $L$ and $T$, i.e. the ones the star would have in a non irradiated evolution. We then let the code solve the stellar structure with these new boundary conditions.

When $L_h$ becomes high enough, either because there is a MSP in the system or we are in a phase with high values of mass transfer, it could happen that the difference in Eq.5 in DE93 (EQ. \ref{EQN_L} here) becomes negative and the code can not integrate the evolution of stellar structure. In these cases, we iteratively increase $T_{int}$ by a few percent (2.5\%) of $T_h$ each iteration, until said difference does not become positive again. In other words, we are allowing the star to absorb part of the energy emitted by the NS to further increase its temperature to be able to emit a non zero luminosity again. 

It will be helpful for the discussion in the following sections to recall the equation for the orbital period derivative in its most general case;
defining $q=M_2/M_1$, and combining the time derivative of the angular momentum ($J$) and the Third Kepler Law we get:

\begin{equation}
\frac{\dot P}{P}= 3\frac{\dot J}{J} - 3 \frac{\dot M_2}{M_2}(1-q\eta_{\rm accr})+\frac{\dot{(M_1+M_2)}}{M_1 + M_2}.
\label{EQN_PDOT}
\end{equation}

\subsection{Model parameters}
\label{MODEL_PARAMS}

We adopt the MLT \citep{bohm-vitense_1958} convective model where we choose a value of $\alpha_{\rm MLT}= 0.5 $. This choice has been made to obtain a better agreement between the radii obtained from our models and the ones measured for the binary system CM Draconis \citep{morales_2009,torres_2010}.

We start from a secondary mass of $0.75\ M_\odot$ and a NS mass of $1.33\ M_\odot$. The initial orbital separation is $2.27\ R_\odot$ which gives us a starting orbital period of $P_b \simeq 6.6\ hr$. We adopt a value of $6\times 10^8 G$ or $3.0 \times 10^{26} G\ cm^3$ for the magnetic field and the magnetic momentum of the NS, respectively, assuming a starting radius of $10km$, with a spin frequency ($\nu$) of $10\,Hz$.
Different evolutionary tracks are computed by assuming different efficiency of the irradiation, $\epsilon$:  0.01, 0.025 and 0.05. We also follow the evolution in the case of no irradiation and two other evolutions having $\epsilon_X=0.01$ and  $\epsilon_{NS}=0.00$ and vice-versa; with the intent to compare how different contributions affect the evolution. We follow these models to the point where the secondary mass reaches a value slightly lower than the minimum value possible for the donor mass of \saxj. In our calculations we adopt a minimum time-step of $10^{3}\ yrs$.

\section{System evolution}
\label{EVO}
In this section we will describe the results. We first deal with the general evolution of the systems and then with the variation of selected physical properties of both the companion and the neutron star. 

\subsection{General evolution}
\label{EVO_GEN}

\begin{figure*}
\vspace{-0.5cm}
\includegraphics[width=1.9\columnwidth]{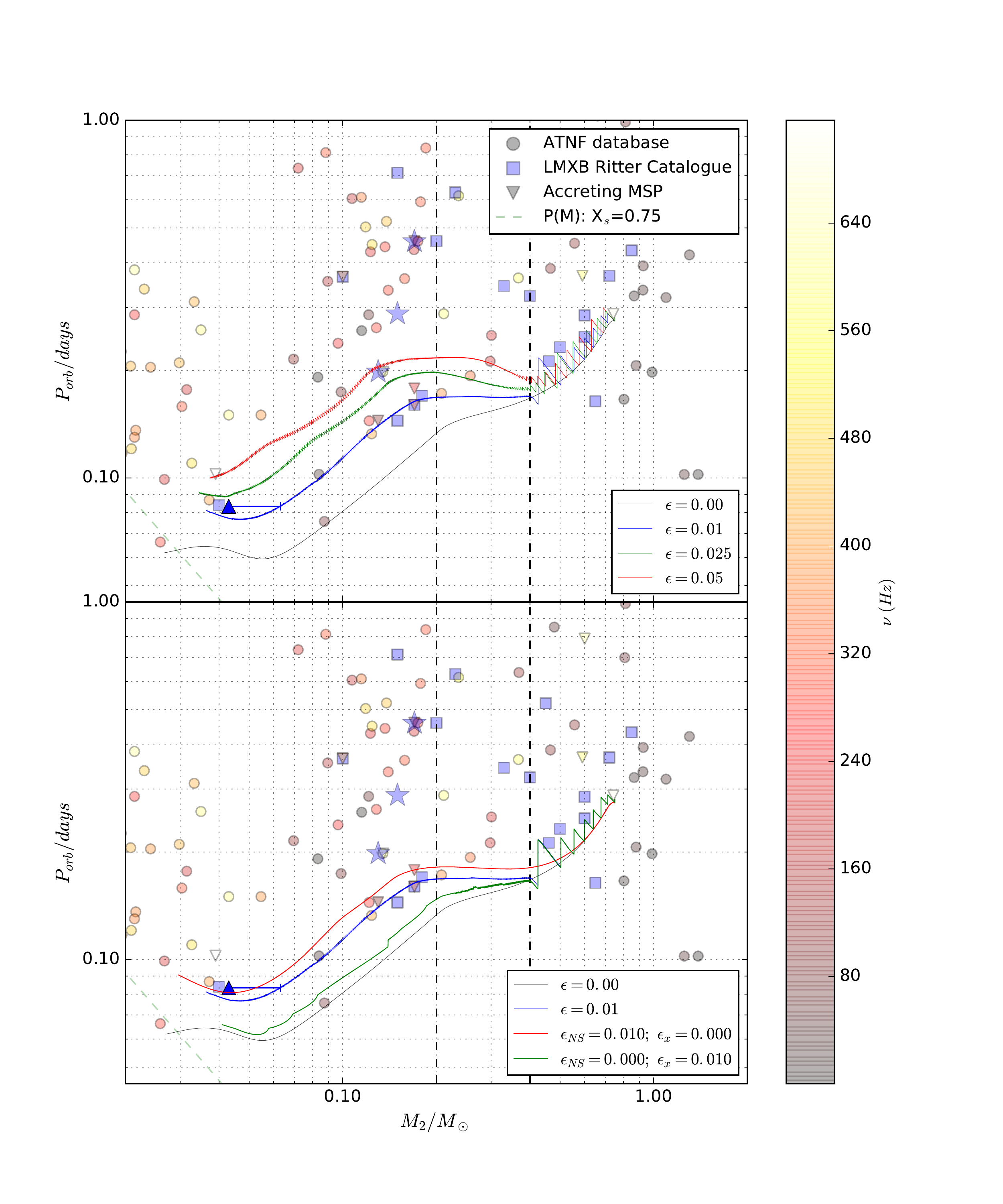}
\vspace{-1.2cm}
\caption{\textit{Upper panel:} The evolution of the models with equal irradiation efficiency on both X-ray and pulsar luminosity contribution in the $M_2$ -- $P_{orb}$ plane. We also report the data of the minimum mass and the orbital period of those system in the ATNF and Ritter catalogue, while also including the position of the three transitional MSPs and the known accreting MSPs . Where available, the dots representing the data are color coded according the spin frequency ($\nu$) of the pulsar in the system. The two blue stars represent the position in this diagram of two transitional MSP, while the solid blue triangle marks the position of \saxj. In highlighting the position of \saxj\ we included the a plausible range of donor mass ($\delta M_2 = 0.02$) for the system. The green dashed line represent the mass -- period relation for a structure with a mass -- radius relation exponent of $-1/3$ and surface hydrogen, $X=0.75$. \textit{Lower panel:} The evolution of some model calculated with different recipes for the irradiation where only one source of energy is considered.}
\label{FIG_MVSP}
\end{figure*}

\begin{figure}
\hspace{-1cm}
\includegraphics[width=1.2\columnwidth]{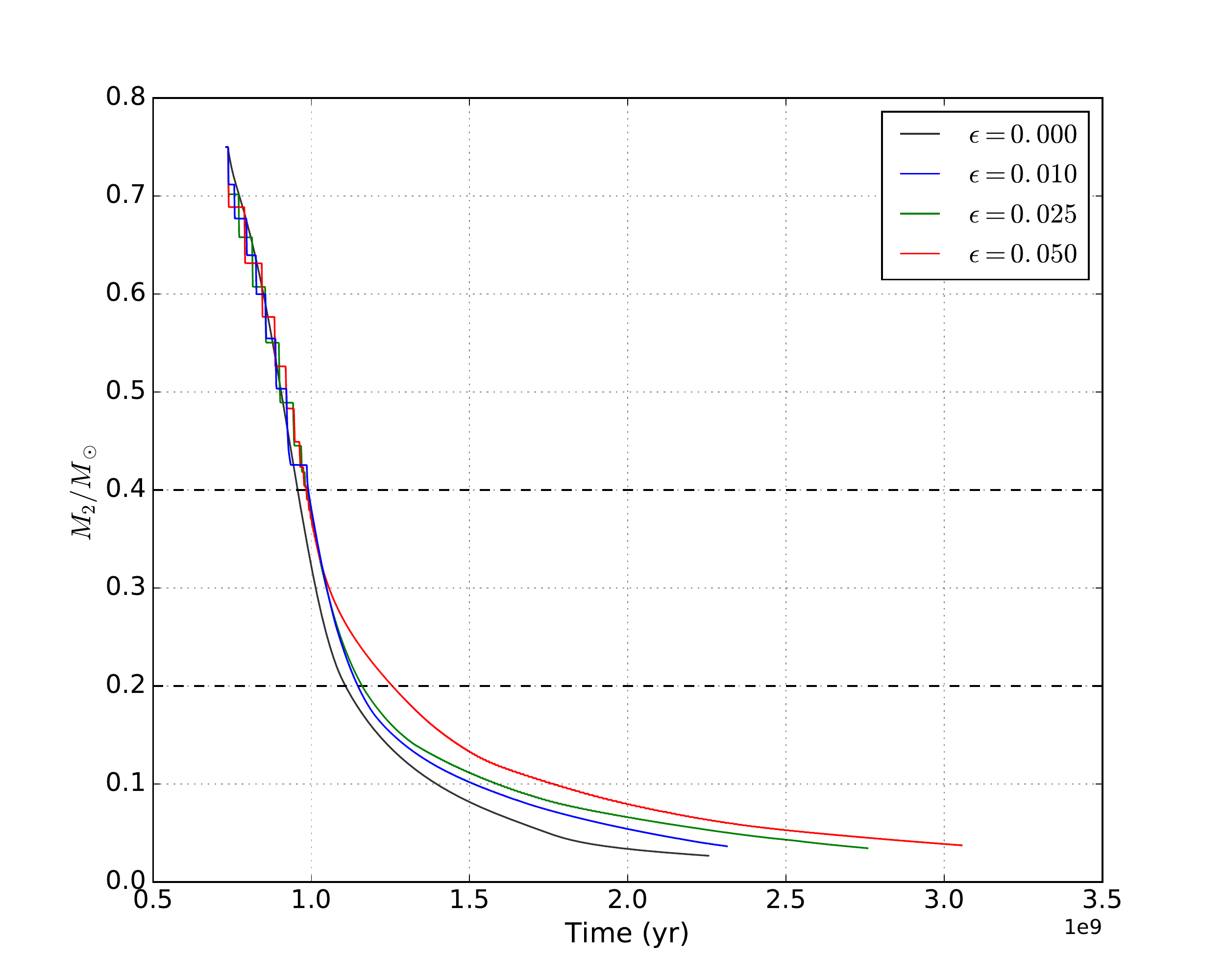}
\vspace{-0.8cm}
\caption{The evolution of the companion mass ($M_2$) with time for the models with efficiency values labelled. The black, dashed lines separate the three phases described in the text. A closer inspection of the tracks reveals that these models spend more than half of their lifetime in the last phase of evolution (see Table \ref{TAB_ratio})}
\label{FIG_TVM}
\end{figure}

\begin{figure}
\hspace{-0.8cm}
\includegraphics[width=1.2\columnwidth]{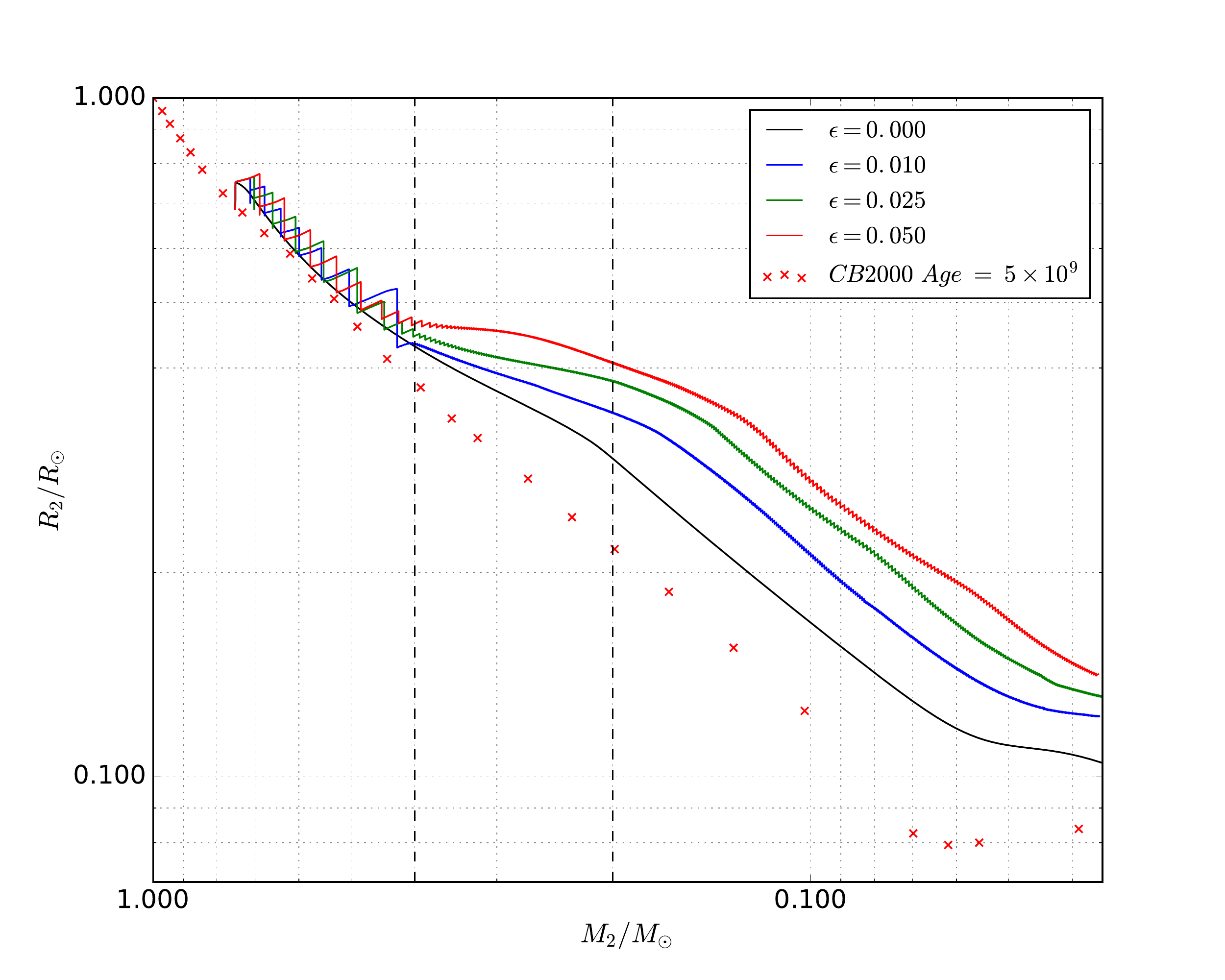}
\vspace{-0.8cm}
\caption{The mass -- radius relation for the three models labelled. As in Figure \ref{FIG_MVSP} and \ref{FIG_TVM} the black, dashed lines separate the evolutionary phases described in the text. The two series of cross represent the mass - radius relations from CB2000 of the labelled age. At first glance, it is evident how, later in the evolution, the models deviate from the equilibrium, even in the non irradiated case.}
\label{FIG_MVR}
\end{figure}

Figure \ref{FIG_MVSP} reports the evolution of the models in the $M_2/M_\odot -P_{orb}$ plane as well as the minimum mass data from two currently available databases, the Ritter LMXB and the ATNF\footnote{ see also http://www.atnf.csiro.au/research/pulsar/psrcat} catalogues \citep{manchester_2005,ritter_2003}. We highlighted the position of \saxj\ in this plane with a solid, blue triangle; while also reporting a reasonable value range for its mass (see \S~\ref{EVO_1808}).  We also included the three transitional millisecond pulsars XSSJ12270-4859 \citep{demartino_2015,papitto_2015}, PSRJ1023+0038 \citep{archibald_2009,patruno_2014,papitto_2015b} and  IGR J18245–2452 \citep{papitto2013}. The figure also highlights the currently known accreting MSP, reported as a reversed triangle.
Where available, the points  are color coded according to the spin of the pulsar in the system, as labelled. The upper panel of the figure describes the behaviour of the models with equal efficiency on both sources of irradiation, while lower panel reports those models calculated with different recipes, as labelled.
When inspecting the upper panel in Figure \ref{FIG_MVSP}, we clearly see that we are able to bracket \saxj\ between two models and draw conclusions on its previous evolution.

We can roughly divide the evolution in three phases, highlighted in the figure by black, dashed lines: an early phase ($M_2/M_\odot > 0.40$) where large ($\sim 1 - 0.5\ hr$) variations in period are present within a cycle, a subsequent phase ($0.40 > M_2/M_\odot > 0.20$) where the period tends either to increase or to remain at a constant value, and a last phase ($M_2/M_\odot < 0.20$) where the period decreases to its minimum value, before increasing again. 
Both the minimum period and the mass at which it is reached vary with the efficiency of the irradiation; in these evolutions the mass at the minimum period is  $0.03< M_2/M_\odot < 0.05$.  Along the whole evolution, the system goes through cycles similar mass loss cycles, as we will see in \S~\ref{EVO_DONOR}.  

From the bottom panel of Figure \ref{FIG_MVSP} we see that the standard period--mass evolution with no irradiation ($\epsilon_x=\epsilon_{NS}=0$) follows a path quite distant from the values measured for \saxj. We further see that the early phases of an irradiated evolution including only irradiation by the pulsar spin down luminosity (i.e $\epsilon_x=0$), is similar to the standard evolution; on the contrary, at later stages, the orbital period remains larger than in the standard case, and similar to the periods obtained by including both irradiation contributions.  
Viceversa, the early evolution including only the X ray irradiation (i.e $\epsilon_{NS}=0$) displays cycles similar to those we see in the evolution including both contributions, while, at later stages, the orbital period versus mass relation becomes similar to the standard case. 

{\it Thus the MSP irradiation contribution is fundamental to reach the \saxj\ stage}.
Indeed, because of the cycles that characterise the entire evolution, the system spends most of its lifetime in a state where mass transfer is not active (see Figure \ref{FIG_MDOT} in \S~\ref{EVO_DONOR}), thus the radiation from the pulsar is the driving mechanism of the long term behaviour of these models, as it keeps the companion star bloated and out of equilibrium, allowing the system to remain at, or reach, larger orbital periods. 
 
Figure \ref{FIG_TVM} describes the evolution of the donor mass with time along the four sequences labelled. A closer inspection reveals that the three phases do not have equal duration, instead, even for a non irradiated model, the systems will spend a larger part of its lifetime in the last of the phases we indicated previously, i.e. $M_2/M_\odot < 0.20$.
 With our models we have then the possibility to understand why the $M_2/M_\odot -P_{orb}$ plane is more populated toward the red back and black widow regions. In fact, it will be more likely to observe a system in the $M_2/M_\odot < 0.20$ range, as the time ratios are $\sim 0.15; \sim 0.10; \sim 0.75$ for each of the three phases, respectively (see Table \ref{TAB_ratio}).

In Figure \ref{FIG_MVR} we plot the models shown in the upper panel of Figure \ref{FIG_MVSP} in the  $M_2/M_\odot - R_2/R_\odot$ plane, where we can again identify the three different  phases. 
The first one $0.75 < M_2/M_\odot < 0.40$ features large, cyclic variations of the radius, which are the counterpart of the cyclic variations in orbital period shown in Figure \ref{FIG_MVSP}. Moreover in this phase the radius tends to return to its non irradiated value between two cycles. 
The other two phases, although not much evident, can indeed be distinguished. In fact, the general steepness of the mass radius relation is different between the two.
In Figure \ref{FIG_MVR} we also plot the mass-radius relation obtained from isolated equilibrium model \citep[][CB2000 as labelled]{chabrier_2000}. The deviation of the mass--radius relation of these evolution from the boundary line of the thermal equilibrium models (crosses in the figure) is progressive with decreasing mass, and also very important for the standard evolution with no irradiation.

We take the opportunity to comment the fact that when calculated with our numerical models, the exponent of the general mass radius relation is never equal -1/3 when the star enters the regime were it should be completely convective \citep{king_1988}. Instead, by looking at Figure \ref{FIG_MVR}, we see that it is always positive, at best close to zero, even for a non irradiated model.

\begin{table}
\centering
\begin{tabular}{cccc}
\hline
\hline
$\epsilon$ & \multicolumn{3}{c}{$M_2/M_\odot$}\\
\hline
& $0.75\div 0.4$ & $0.4\div 0.2$ & $0.2\div 0.02$ \\
\hline
0.010 & 0.17 & 0.11 & 0.72\\
0.025 & 0.12 & 0.10 & 0.78\\
0.050 & 0.12 & 0.14 & 0.74\\
\hline
\hline
\end{tabular}
\caption{The time spent (expressed as ratio over the total model life time) in each of the three phases identified in \S~\ref{EVO_GEN} for each of the three models. }
\label{TAB_ratio}
\end{table}

\subsection{Donor evolution}
\label{EVO_DONOR}

\begin{figure*}
\vspace{-0.5cm}
\includegraphics[width=2.0\columnwidth]{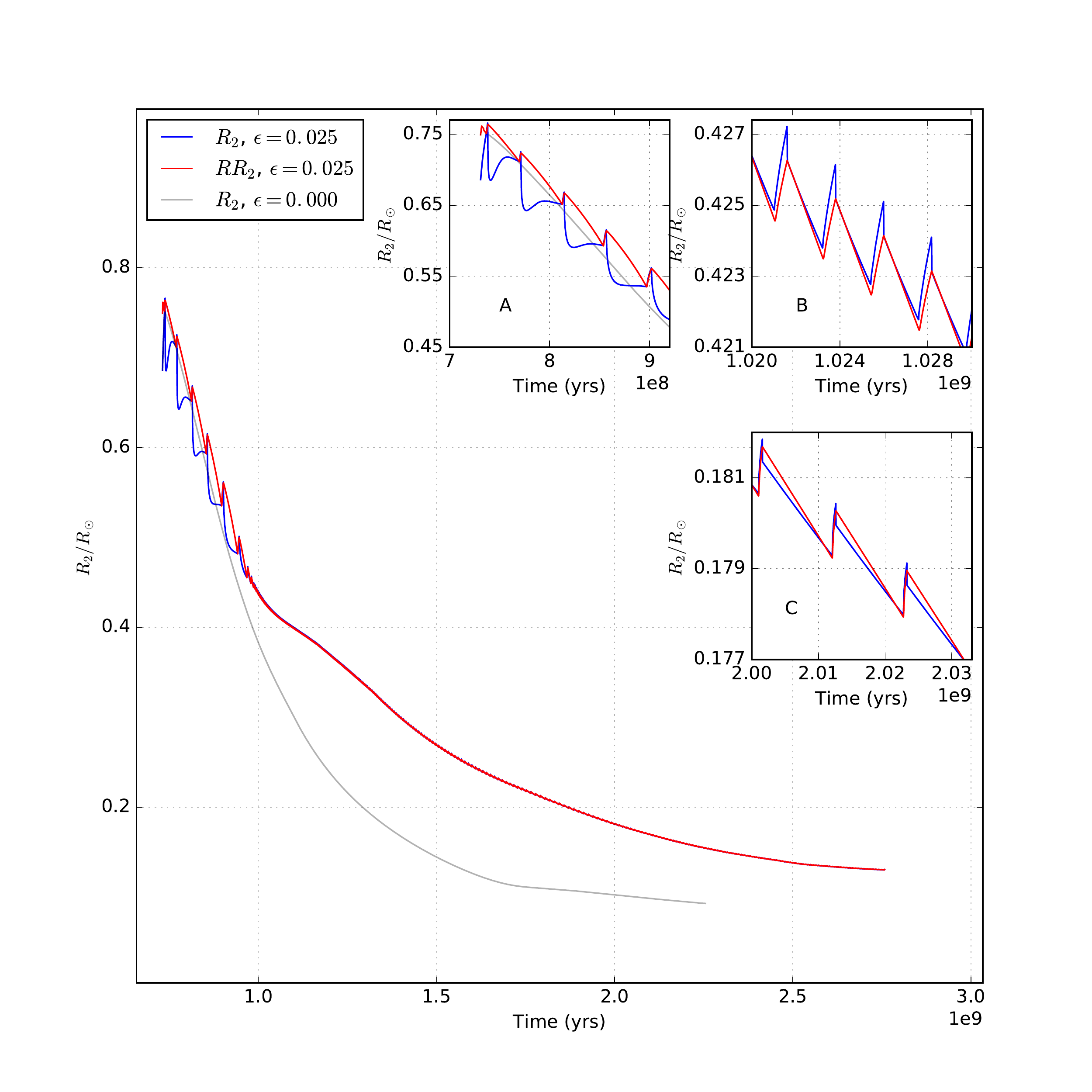}
\vspace{-1cm}
\caption{\textit{Main panel:} The evolution of the companion star radius ($R_2$, solid, blue line) of the model with $\epsilon = 0.025$, as function of time, compared with its non irradiated counterpart (black, line). We also report the evolution of the companion Roche lobe radius for the irradiated model ($RR_2$, solid, red line). \textit{Panel A, B, C:} Three magnified views of the radius evolution corresponding to different ages. More specifically between $0.700 - 0.920\ Gyr$; $1.02 - 1.03\ Gyr$; $2.00 - 2.033\ Gyr$ for Panel A, B and C respectively. The tracks are color coded as in the main panel. }
\label{FIG_TVR_RR}
\end{figure*}

\begin{figure}
\hspace{-0.8cm}
\includegraphics[width=1.2\columnwidth]{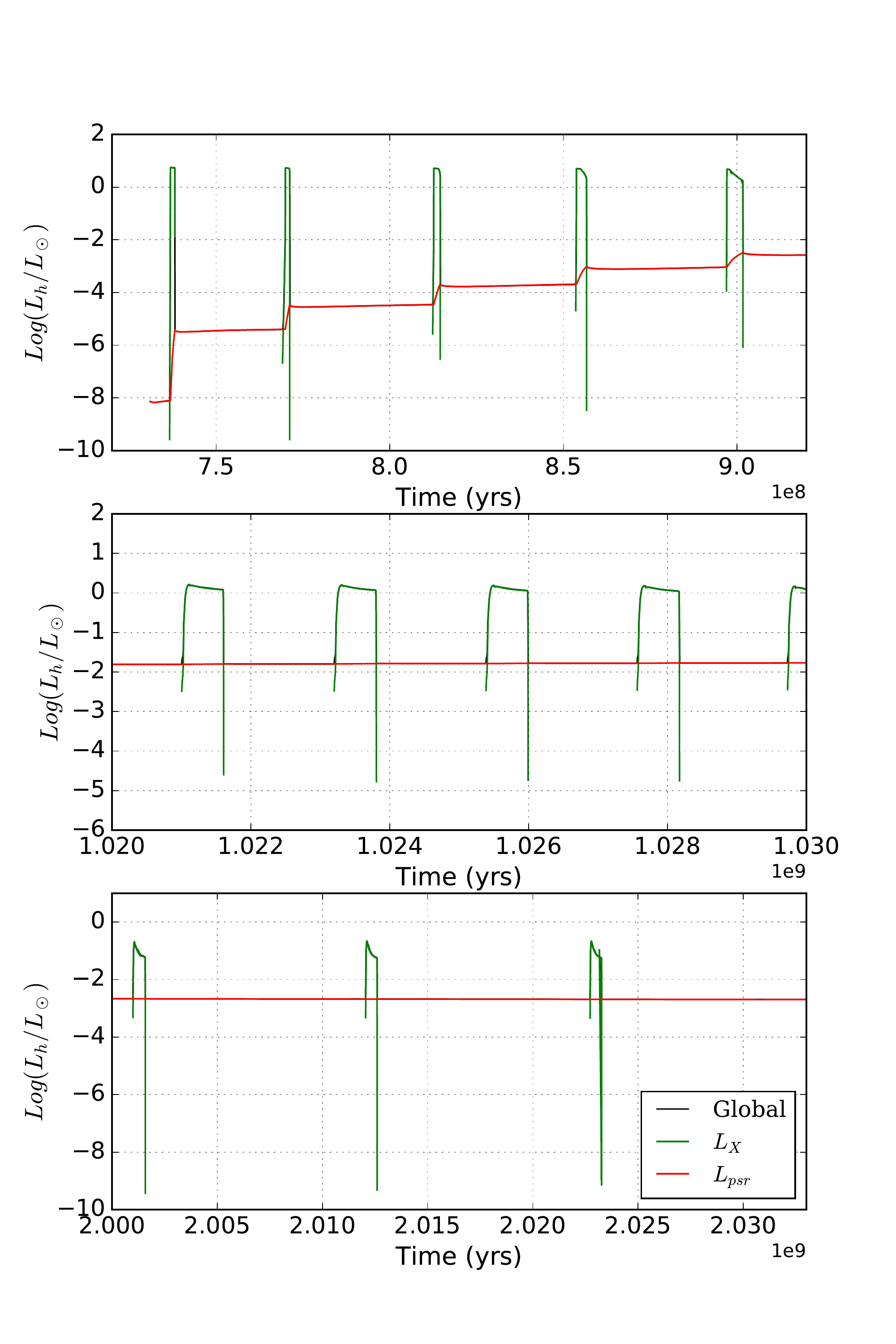}
\vspace{-1cm}
\caption{The total heating luminosity on the companion star ($L_H$) during the three phases shown in panel A,B,C of Figure \ref{FIG_TVR_RR} (upper, middle and bottom panel, respectively). We colour coded the global value of $L_h$ and the contributions of both X-ray luminosity ($L_X$) and the pulsar luminosity ($L_{psr}$), as labelled. }
\label{FIG_LH}
\end{figure}

\begin{figure}
\hspace{-0.8cm}
\includegraphics[width=1.2\columnwidth]{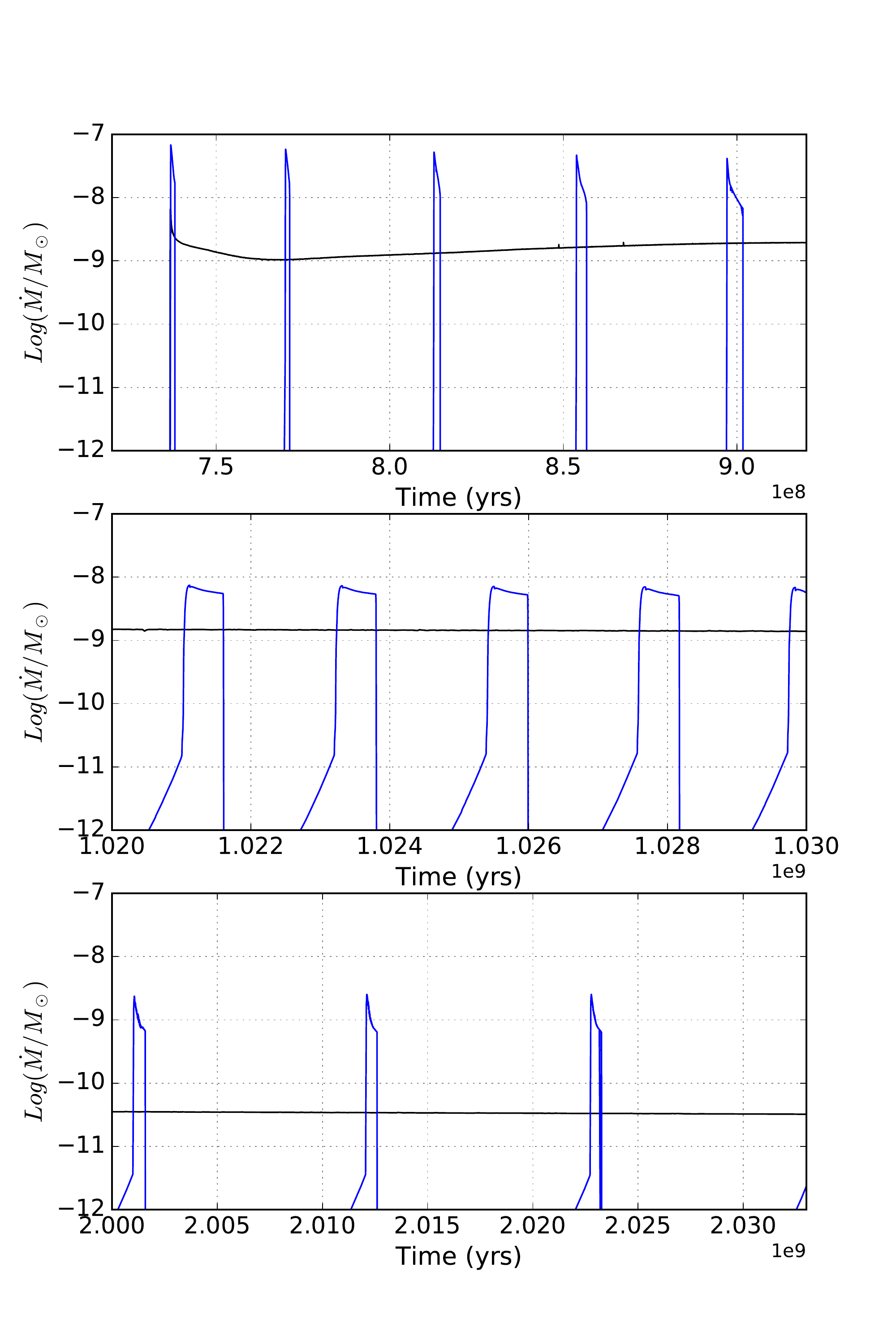}
\vspace{-1cm}
\caption{The evolution of the mass lost by the companion (as $Log(M_2/M_\odot)$) during the evolution segments reported in Panels A, B and C of Figure \ref{FIG_MVSP}, respectively in the upper, middle and bottom panels. Each panel highlights the cyclic nature of the irradiated evolution throughout its entire duration.}
\label{FIG_MDOT}
\end{figure}

\begin{figure}
\hspace{-0.8cm}
\includegraphics[width=1.2\columnwidth]{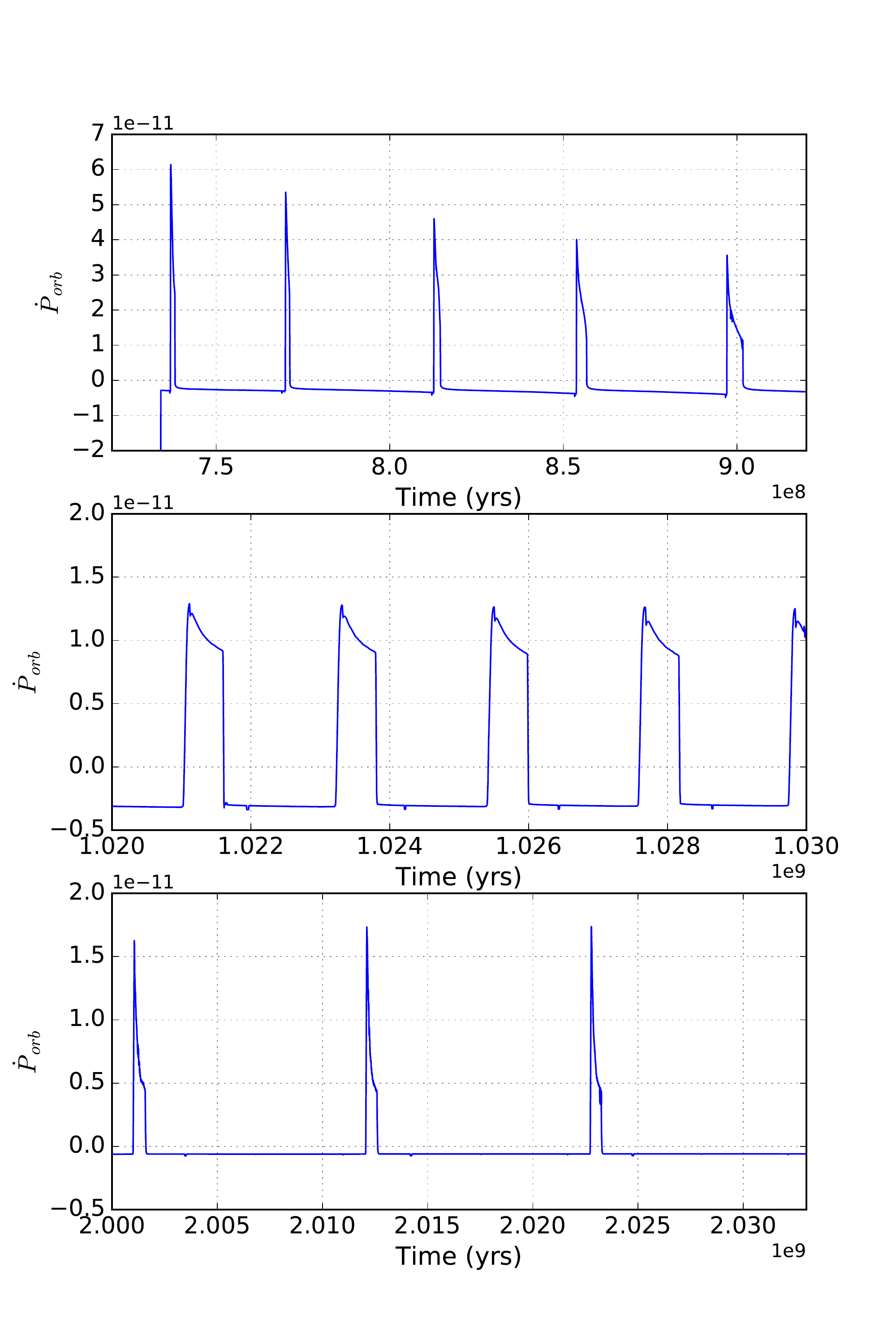}
\vspace{-1cm}
\caption{The values of the $\dot P_{orb}$ in the segments of the model identified in Panels A, B and C in Figure \ref{FIG_TVR_RR}. As described in the text, the high value of $\dot P$ during the mass transfer phases of the cycles is due to the feedback mechanism described in \S~\ref{EVO_DONOR}}
\label{FIG_TVPDOT}
\end{figure}

\begin{figure}
\hspace{-0.8cm}
\includegraphics[width=1.2\columnwidth]{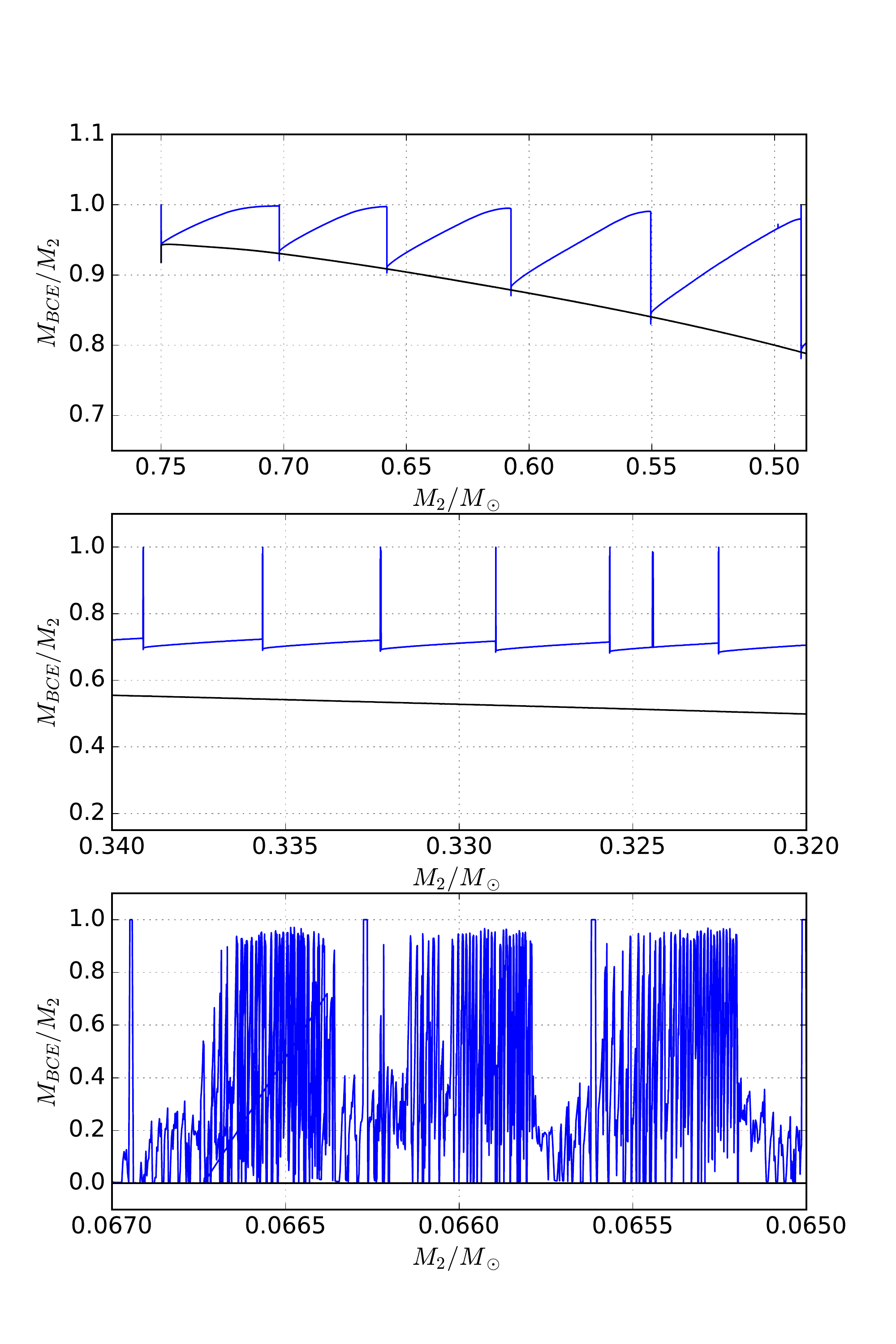}
\vspace{-1cm}
\caption{The location, in mass ratio, of the convective envelope ($M_{BCE}$) in the part of evolution described in Panel A,B and C of Figure \ref{FIG_TVR_RR} (upper, middle and bottom panel, respectively). We also plot the value of the same quantity for a non irradiated model (black, solid line). The different behaviour between the two models is indeed crucial when describing the evolution of such models toward the black widow region (see text). }
\label{FIG_MVMBCE}
\end{figure}

In order to keep the figures of this section as readable as possible, we will discuss the evolution of the system calculated with $\epsilon =0.025$ noting that the same behaviours can be observed in all three models.

Fig. \ref{FIG_TVR_RR} describes the evolution of both the donor radius ($R_2$, blue) and its Roche lobe radius ($RR_2$, red) as function of time. In the figure we also plot the radius of the non-irradiated model (black). The three panels (A,B,C) describe snapshots of the evolution belonging to each one of the three phases we identified in \S~\ref{EVO_GEN}. 
Keeping this separation in mind, Figures  \ref{FIG_LH} to \ref{FIG_MVMBCE} describe the evolution of some selected properties of the companion star and the system in the same sections of the model. The evolution of these three evolutionary segments are described in the upper, middle and lower panel in each figure, respectively. 
We plot in Figure \ref{FIG_LH} the total heating luminosity on the companion star; in the figure we colour coded the contribution of the X radiation ($L_{\rm x}$), when non zero, as the solid, green line, and the pulsar luminosity  ($L_{\rm psr}$) as the solid, red one, while the global contribution is represented as a black solid line.  
Figure \ref{FIG_MDOT} and \ref{FIG_TVPDOT} describe the evolution of  $\dot M$  and $\dot P$ in the three segments in Figure \ref{FIG_TVR_RR} as function of time, respectively. Figure \ref{FIG_MVMBCE}, on the other hand, describes, as a function of the companion mass, the evolution of the outer convective envelope of the donor during the stages of the evolution we identified.  We plot the mass ratio of the position where the inner boundary of the convective envelope is located. With this representation a value of zero describes a fully convective star. We note that here the values of mass correspond to the same section described in Figures \ref{FIG_LH} to \ref{FIG_TVPDOT}.  

The main feature we notice from Figure \ref{FIG_TVR_RR} is that the evolution has a cyclic nature during its entire length. Moreover, as panels A, B and C show, the cycles are different in shape, amplitude, and duration. \\

Panel A refers to the first part of the evolution, more specifically to the first $10^9$ years. Here the companion star has a mass of $0.75 \lesssim M_2/M_\odot \lesssim 0.50$. The cycles we observe in the model during this first phase are similar to the ones described in works from other authors \citep[][and many others]{pod_1991,benvenuto_2014,benvenuto_2017} in models of similar mass. 
Each cycle lasts for few $10^7 yr$ and it is made up by two phases: a short period of time (few Myr) were the system transfers mass from the companion to the neutron star and a longer one where, once it has detached, the companion star radius shrinks toward the thermal equilibrium radius. In this phase, the system spends most of its lifetime in a detached state (as Figures \ref{FIG_LH} to \ref{FIG_TVPDOT} show). 
When mass transfer is active it is very efficient, so that the mass lost by the companion reaches values greater than the Eddington limit\footnote{We take the opportunity to remind to the reader that the system looses this excess mass and its angular momentum accordingly}. The irradiation luminosity ($L_h$) during the first few cycles is about ten to five times $L_\odot$.
It is worth noting that here the main contribution to $L_h$ is given mostly by the X radiation as the luminosity of the still slow pulsar is many orders of magnitude lower, as the plot in the upper panel of Figure \ref{FIG_LH} shows. Indeed, when the system is detached the star basically behave as if it is not irradiated, as $L_h$ is so low that it is negligible.
The X--ray luminosity blanket leads to the expansion of the companion star on the the thermal time scale of its convective envelope \citep[as described by DE93 and][]{dantona1996}. Qualitatively, equation 14.14 in \cite{dantona1996}, written below, describes the behaviour of the donor radius:

\begin{equation}
\left[ \frac{\delta ln(R_2)}{\delta t}\right]= \left[ \frac{\delta ln(R_2)}{\delta t}\right]_{\dot M=0} + \left[ \frac{\delta ln(R_2)}{\delta t}\right]_{ill}
\label{EQN_DR}
\end{equation}

The first term is the thermal relaxation one, which, for stars with convective envelopes such as our donors, is negative, and acts to limit the mass transfer. The second term is the radius expansion due to irradiation. This represents the reaction of the star to the locking of a part of its surface by the external source.  The radius expansion then enhances mass loss and consequently the X-ray irradiation, in a feedback effect responsible for the period increase during the mass transfer phase of each cycle along the entire evolution  (see Equation \ref{EQN_PDOT}), as shown in Figure \ref{FIG_TVPDOT}. We also comment on the fact that the star, when irradiated, here and in the next evolutionary phases, is slightly larger than its Roche lobe radius.

The very high peaks of mass loss, as seen in Figure \ref{FIG_MDOT}, are due, then, to the fast reaction of the star, which, at these masses, has a relatively small convective envelope and thus  expands on a short time scale. Consequently, mass loss dominates the period evolution and the orbital period, separation and Roche lobe radius increase (see Eq. \ref{EQN_PDOT} and Figure \ref{FIG_TVPDOT}).
When the thermal relaxation of the envelope takes over, the star radius shrinks and the system detaches. Eventually, radio ejection activates interrupting the accretion of mass onto the primary and drastically reducing the X-ray contribution to irradiation. Indeed, as we see from Equation \ref{EQN_REJ}, $P_{\rm crit}$ is proportional to the luminosity of the system ($L_{\rm 36}$). A new cycle can follow only when the angular momentum losses bring the donor in contact\footnote{As noted before this happens when $((R_2-RR_2)/Hp)\simeq 0 $, because in \cite{ritter_1988} scheme we do not impose a strict contact criterion.} and mass loss becomes large enough to overcome the radio ejection mechanism (i.e. $P_{\rm crit} > P_ {\rm orb}$) thus allowing accretion onto the primary again.
An interesting behaviour that emerges from our models is that, in this first portion of evolution, the phase where the system is detached in each cycle is so long (see Figure \ref{FIG_TVR_RR},\ref{FIG_LH} and  \ref{FIG_TVPDOT})  the star starts increasing its radius again because of its nuclear evolution. This is even more evident in these first four cycles where the Kelvin Helmoltz time scale is shorter.
The irradiation phenomenon has another effect on the internal structure of the companion star that will play a primary role in shaping its evolution in later stages. Looking at the upper panel of Figure \ref{FIG_MVMBCE} we see that in this first phases the value of the mass where the bottom of the convective envelope is located ($M_{\rm BCE}$) is shifted to higher values during the phases where mass transfer is active, especially if we compare the position of this layer to a non irradiated model. In this early stage, this effect is of little relevance as the star has an already small convective envelope; but when it will approach the red back and the black widow regime it will play a crucial role in shaping the path the system will follow (see later).\\

Panel B in Figure \ref{FIG_TVR_RR} represents a later stage of the evolution in the model, between $1.02$ and $1.03$ Gyr; here the companion mass is $0.34 \lesssim M_2/M_\odot \lesssim 0.32$. 
The cycles in this phases are shorter, as a single one lasts for few Myr. The comparison with Panel A shows that the cycles have here a completely different shape. In this second phase the orbital period tends to increase or to remain at a constant value.
This happens because of a few factors. 
Because of the increased depth of the convective envelope for these smaller donor masses, the radius expansion and contraction occur on a longer time-scale; consequentially the system, as radio ejection activates, does not becomes completely detached between two cycles, i.e. $\dot M$ is not rigorously zero\footnote{It is however not enough to overcome radio ejection}.  Also, because the pulsar is now quite fast\footnote{In this specific case the neutron star is rotating at about $550 Hz$, as we are showing a portion near its frequency peak, see Figure \ref{FIG_MVNS} and \S~\ref{EVO_NS}} its contribution to the irradiation energy is no more negligible, when the system is detached, thus there is a persistent irradiation on the donor (see middle panel of Figure \ref{FIG_LH}).
The net effect is to keep the companion star bloated for the entire duration of this phase, as the term due to irradiation in Equation \ref{EQN_DR} is always non-zero\footnote{Although it may not be high enough to overcome the thermal relaxation term in some parts.}; favouring the increase in orbital period.
Moreover, being the pulsar now a MSP, the radio ejection mechanism \citep{burderi_2001} has a more important role, thus a higher value of $\dot M$ is required to start a new cycle ($|\dot M/(M_\odot\ yr)| \gtrsim 10^{-11}$ as Figure \ref{FIG_MDOT} and \ref{FIG_TVPDOT} show).
Middle panel in Figure \ref{FIG_MVMBCE} shows that, as in the previous case, the model has a less extended convective envelope than the in non irradiated case. Contrary to the first case of evolution (upper panel), due to the pulsar irradiation, the convective extension does never increase up to that of the non irradiated model, when mass transfer is not active. \\

Panel C in Figure \ref{FIG_TVR_RR} represents a stage of the evolution in the model corresponding to ages between $2.00$ and $2.033$ Gyr. Here the companion mass is $0.065 \lesssim M_2/M_\odot \lesssim 0.067$. The shape of the cycles in Figure \ref{FIG_TVR_RR} is different from the two previous phases, and the system spends most of its time in the detached phase, thus the orbital period is globally decreasing again. 
The duration of the cycles in this portion of evolution is about $10 Myr$ (for the $\epsilon = 0.025$ model), twice as long than the cycles in the previous part. In spite of the complete cycle being longer, the phase where mass loss is active is shorter, as we see comparing both middle and bottom panels in Figure \ref{FIG_LH}. This happens because the pulsar, being a MSP, is so fast that the mechanism of radio ejection becomes more efficient and stops accretion even before it stops due to the stellar envelope thermal relaxation. 
When radio-ejection is active, the X luminosity drops, and the secondary star radius goes on decreasing  due to thermal relaxation. This negative feedback is the counterpart of the positive feedback producing the period increase when irradiation is effective. Although the period again increases during the peak mass transfer phases, the angular momentum losses, from both MB and GW, dominate the evolution (see bottom  panel of Figures \ref{FIG_LH} to \ref{FIG_TVPDOT}). Global irradiation is the leading factor determining the structure of the companion star, whose mass, in the absence of this external source of energy, would have the structure of a plain brown dwarf, while it still remains not fully convective (bottom panel in Figure \ref{FIG_MVMBCE}), justifying our choice of keeping the magnetic braking active.

\subsection{Pulsar evolution}
\label{EVO_NS}

\begin{figure}
\hspace{-0.8cm}
\includegraphics[width=1.2\columnwidth]{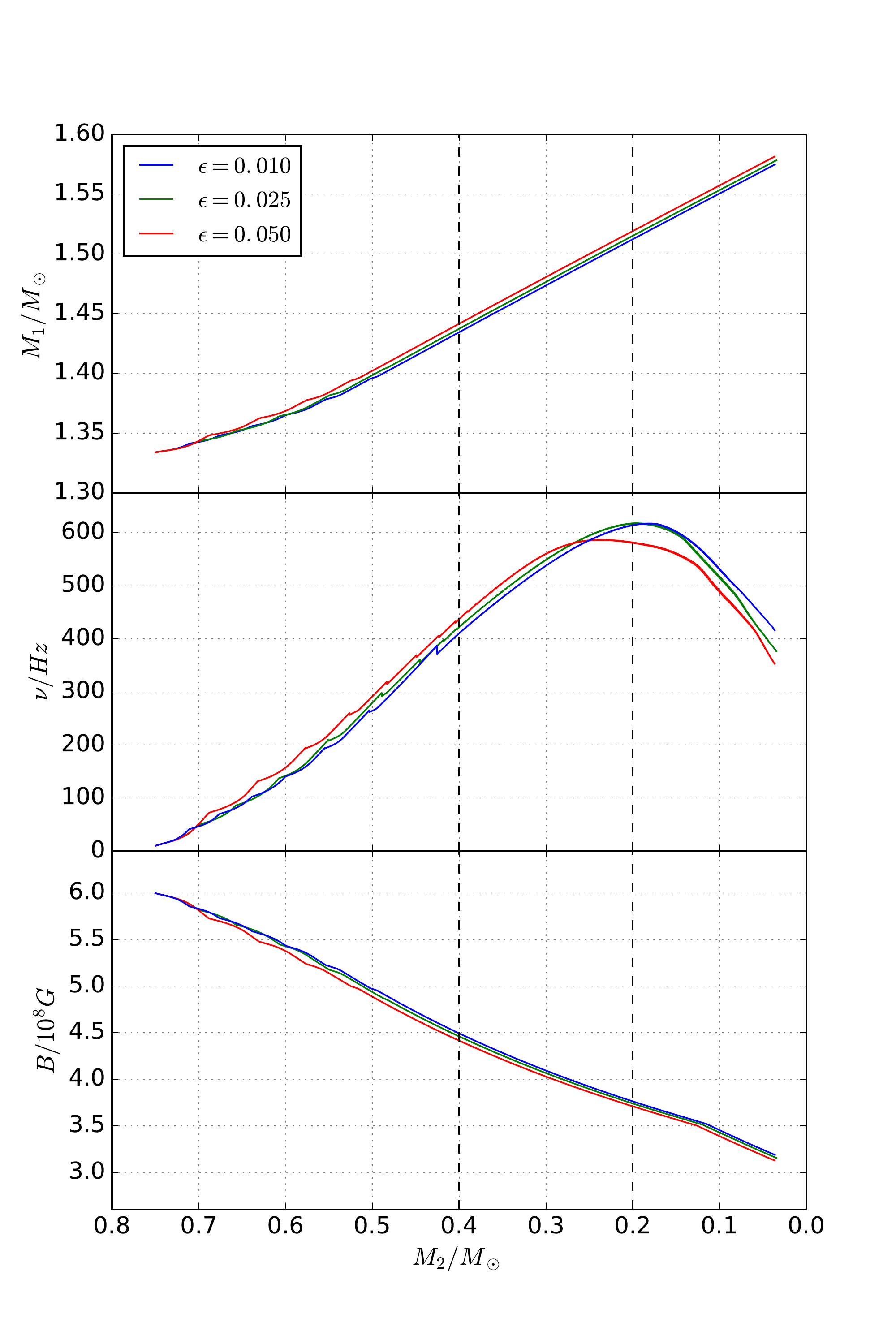}
\vspace{-1cm}
\caption{NS mass (top panel), spin frequency (middle panel) and NS magnetic field ($B$), in units of $10^8G$ (bottom panel) as a function of the donor mass, along the evolutions described in \S~\ref{EVO}.The black dashed lines separate the three evolution phases identified in the text.}
\label{FIG_MVNS}
\end{figure}

\begin{figure}
\hspace{-0.8cm}
\includegraphics[width=1.2\columnwidth]{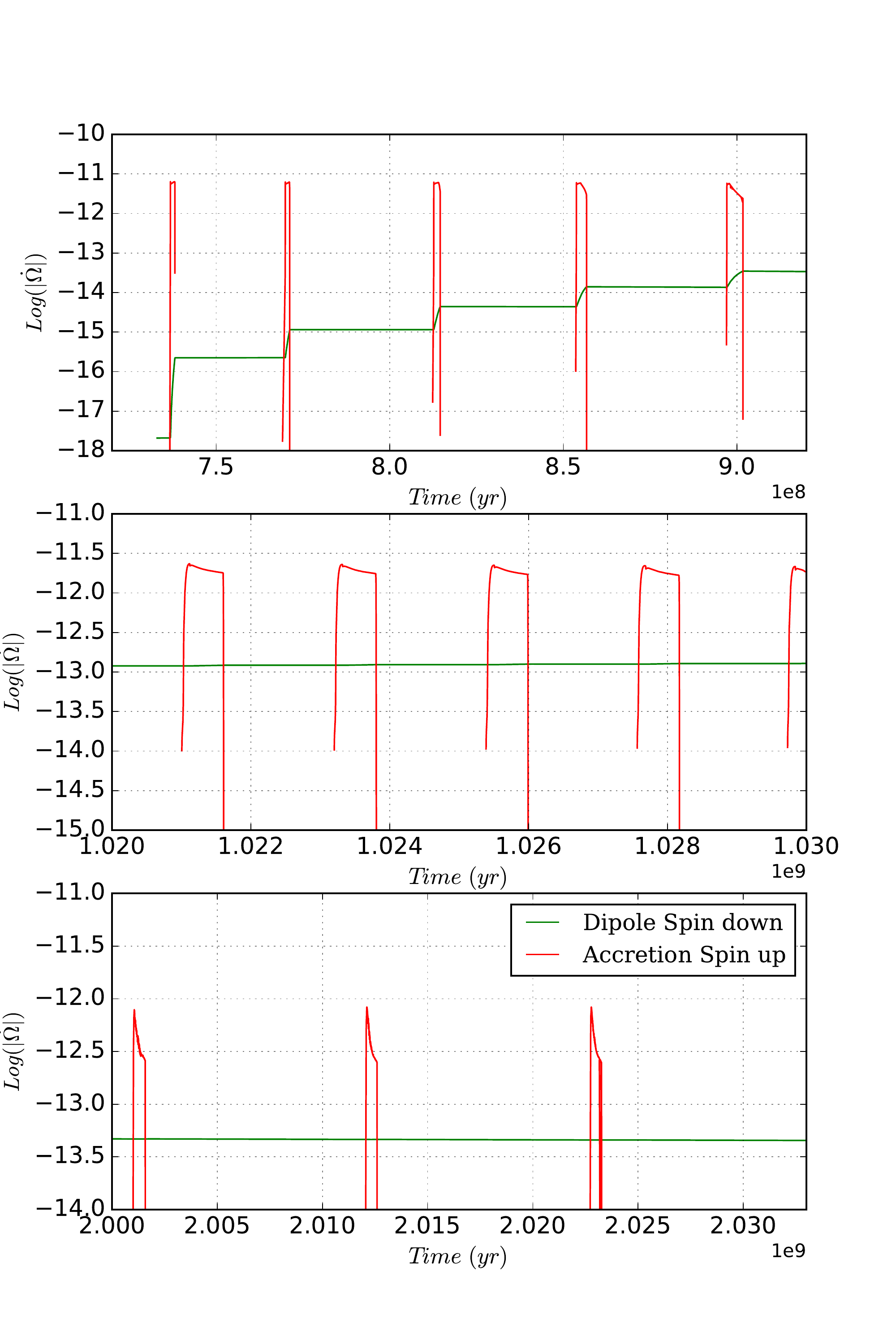}
\vspace{-1cm}
\caption{The comparison between the absolute value of $\dot \Omega$ from both the accretion induced spin up and the dipole spin down. We report these value for the model with $\epsilon =0.025$ in the same three portion of evolution describe in Panels A,B, C of Figure \ref{FIG_TVR_RR}. }
\label{FIG_MVSPINUD}
\end{figure}

Figure \ref{FIG_MVNS} describes the evolution of the NS, primary component of the system, for the secular evolutions previously discussed in terms of the donor component. As a function of the donor mass, the three panels describe the evolution of the NS mass ($M_{\rm 1}$, upper panel), of its spin frequency ($\nu$, middle panel) and of the magnetic field ($B$). Black dashed lines separate the three stages of evolution previously identified.

The NS mass increases, as result of accretion. The final value of $M_{\rm 1}$ in the last models is about $1.57 M_\odot$. Similarly, the magnetic field intensity $B$\ decreases according model \cite{shiba_1989}  (see \S~\ref{MODEL_IRR}),  from the initial value of $B=6\times 10^8\ G$ to a value of $B_{\rm fin}\sim 3.2\times 10^8\ G$. 

On the other hand, $\nu$ has a non monotonic behaviour. Figure \ref{FIG_MVSPINUD} reports the value of $|\dot \Omega|$ for the model with $\epsilon = 0.025$ in the three stages described in in \S~\ref{EVO_DONOR} and in Panels A,B and C of Figure \ref{FIG_TVR_RR}. The two contribution, the spin up from accreted matter (red, when non zero) and the spin down from dipole radiation (green), are plotted as separate tracks in the figure. 

While evolving through the first stage, the high value of mass transfer provokes an efficient spin up of the NS;  further, although the system is detached most of the time, the spin down process is still so inefficient\footnote{Indeed $\dot \Omega_{\rm dip} \varpropto \Omega^3$ as in Equation \ref{EQN_PDIP} }, as $\log(|\dot \Omega|)_{\rm sd,max} \leq -14.0$, that is not able to slow down the pulsar before the next cycle;  the net effect is that the pulsar accelerates.

During the second phase ($0.40\leq M_2/M_\odot \leq 0.20$), the detached periods are short, so that the pulsar keeps spinning up, even if the spin down efficiency is high (see the middle panel in Figure \ref{FIG_MVSPINUD}).

During the third phase ($0.20\leq M_2/M_\odot \leq 0.02$), cycles are longer, with shorter mass transfer stages and lower peaks of mass transfer. Moreover since the NS has already been accelerated to millisecond pulsar, the radio ejection phenomenon is very efficient in limiting the quantity of mass accreted onto the NS; thus the pulsar spins down almost constantly as the contribution of the mass transfer is negligible in the global evolution (see central panel of Figure \ref{FIG_MVNS}).

\subsection{The possible evolution to the present \saxj\ system}
\label{EVO_1808}

\begin{table*}
\centering
\begin{tabular}{ccccccc}
\hline
\hline
&	\multicolumn{2}{c}{$M_2=0.04\ M_\odot$ }& \multicolumn{2}{c}{$M_2=0.06\ M_\odot$ } &  { ~~~\saxj~~~  } \\
\hline
 $\epsilon$ & 0.01 & 0.025& 0.01 & 0.025 & \\
\hline 
$P_{\rm orb} (hr)$& 1.89& 2.14 & 1.95 & 2.44 & 2.01 \\
$M_{\rm psr} (M_\odot)$& 1.57 & 1.58 & 1.56 & 1.57 & \\
$B_{\rm psr}(10^8\ G)$&3.20 &3.17 & 3.28 & 3.26                  &  $2\div 4$\\ 
$\nu_{\rm psr} (Hz)$ & 424.01 & 388.24 & 460.60 & 430.36         &400.98\\  \\
$\dot M_{\rm peak} (M_\odot/yr)$& $3.22\times 10^{-9}$&$4.31\times 10^{-10}$& $3.15\times 10^{-9}$&$2.50\times 10^{-9}$  & $1.2 \div 2\times 10^{-9}$ (during outburst)\\
$\dot M_{\rm mean} (M_\odot/yr)$& $7.77\times 10^{-11}$&$8.27\times 10^{-11}$& $9.98\times 10^{-11}$&$1.14\times 10^{-10}$.           &\\ \\
$\dot P_{\rm orb,peak} (s/s)$& $2.68 \times 10^{-11}$& $4.02 \times 10^{-12}$& $1.79 \times 10^{-11}$& $1.52\times 10^{-12}$    & $3.6 \div 3.8 \times 10^{-12}$ (during outburst)       \\
$\dot P_{\rm orb,mean} (s/s)$&  $3.13 \times 10^{-13}$& $4.88 \times 10^{-13}$&  $2.12 \times 10^{-13}$& $5.83 \times 10^{-13}$         &\\ \\
$\dot \nu_{\rm peak} (Hz/s)$ & $7.66 \times 10^{-13}$& $1.04 \times 10^{-12}$& $8.24 \times 10^{-13}$& $1.01 \times 10^{-12}$ & $1.1 \times 10^{-13}$ (during outburst)  \\
$\dot \nu_{\rm mean} (Hz/s)$  & $ -1.69 \times 10^{-15}$ &  $-1.10 \times 10^{-15}$& $-2.82 \times 10^{-15}$& $-1.96 \times 10^{-15}$             &$-1.5 \times 10^{-15}$ (secular) &\\
\hline
\hline
\end{tabular}
\caption{The values of the listed physical quantities as obtained from our models, at the value of mass corresponding to the minimum and median mass for \saxj. The subscripts \textit{peak} and \textit{mean} refer to the peak of nearest mass loss cycle and the mean value over a few cycles, centred around $M_2=0.04\ M_\odot$ and $M_2=0.06\ M_\odot$ , respectively. }
\label{TAB_1808_MOD}
\end{table*}

From Figure \ref{FIG_MVSP} we see that the location of \saxj\ in the $M_2 - P_{\rm orb}$ plane is bracketed by two of our evolutionary paths and more specifically the $\epsilon = 0.01$ and  $0.025$ ones. In addition, we show in Table  \ref{TAB_1808_MOD} that  most of the relevant parameters measured for the system are also bracketed by these models.
We list the values obtained both for the minimum donor mass in the literature ($M_2=0.04 M_\odot$) and for a larger value ($M_2=0.06 M_\odot$) which represents a reasonable upper limit, if we adopt the inclination given by \cite{papitto_2009,sanna_2017}, Di Salvo (2018, submitted) and the observed mass function \citep{chakrabarty_1998}.

The parameters we obtain for the NS in the system reproduce the observed values. Indeed, the value for the magnetic field is in agreement with the observed value in \cite{cackett_2009}  and  \cite{papitto_2009} ($B = 2\div 4\times\ 10^8\ G$), while only slightly larger than what have been measured in \cite{sanna_2017} and \cite{burderi_2006}\footnote{Albeit this is not their primary result.} ( $B \sim 2 \times\ 10^8\ G$ and $B = 2\div 3 \times\ 10^8\ G$ respectively). 
The values we obtain for the pulsar spin, although slightly larger than the value observed \citep[$\sim 401\ Hz$,][and reference therein]{chakrabarty_1998, sanna_2017}, are still a good match. As we remarked, the standard model not including irradiation, even when maintaining the MB AML active, evolves through much shorter orbital periods at the donor mass, and the same occurs for the model having only X--ray irradiation, because the evolution in this range of masses is dominated by the detached phases (see Figure \ref{FIG_LH} in  \S~\ref{EVO_DONOR}). 
We stress again the role of the persistent irradiation field provided by the spin down radiation of the pulsar, as, keeping the companion star bloated when mass transfer is not active, it is essential to allow maintaining the period in the range observed for \saxj.  

The minimum time step in our simulation is $10^3\ yrs$, so the cycles we observe are a different phenomenon compared to the outbursts observed since the discovery of this system. Thus in one of our time steps there could be thousands of bursts, which we are not able to model within our scheme. Nevertheless, the comparison can be helpful to interpret the long term state of this source. 
Comparing the values of $\dot M_2,\ \dot P_{\rm orb}$ and $\dot \nu$\ to observations (Table \ref{TAB_1808_MOD}) we draw some interesting conclusions.  
We compare the model values to those obtained by \cite{burderi_2006,burderi_2009,hartman_2009} and \cite{sanna_2017} for the 2015, 2008 and 2002 outburst: $\dot M_{\rm 2,b} \sim 2\times 10^{-9}\ M_\odot/yr; \dot P_{\rm orb,b} \sim 3.6\times 10^{-12}\ s/s; \dot \nu_{\rm S} \sim -1.5 \times\ 10^{-15}\ Hz/s $ and $\dot \nu_{\rm b} \ge 1.1\times 10^{-13}\ Hz/s$, where the subscripts \textit{S} and \textit{b} stand for \textit{secular} and \textit{burst}. 
We extract from the model two values for each of these three physical quantities: a peak and a mean value (see Table \ref{TAB_1808_MOD}). The former refers to the peak of the nearest cycle, while the latter is the mean value obtained over few cycles, centred at the value of mass we choose.   

We see that our peak values for $\dot M_2,\ \dot P_{\rm orb}$ and $\dot \nu_{\rm psr}$ are definitely higher than those observed, during a single outburst. From this point of view the values observed could hint the position of \saxj\ on the cycle, indicating that the system, in its long term evolution, is still not at the peak, possibly in the ascending part. We stress that this comparison has to be taken with a grain of salt, as a more detailed modelling and simulations are needed to understand what happens on short-time during a cycle.
  
\section{Discussion and conclusions}
\label{Disc}

We present a series of newly calculated LMXB models with the aim to describe the long term evolutionary path of \saxj, taking into consideration the irradiation phenomenon as primary drive for the evolution. The irradiation model takes into account both the contribution of the X-ray radiation originating from mass transfer, and the luminosity of the pulsar itself. The scheme we follow to implement irradiation, once the total irradiation energy has been calculated, is a modification of the one described in DE93 (as described in \S~\ref{MODELS}). 

The models bracket the position of \saxj\ in the $M_2$ -- $P_{orb}$ plane (Figure \ref{FIG_MVSP}) and studying the $\epsilon = 0.01$ and $0.025$ ones we can draw conclusions on its past evolution. Indeed we have models that go trough thousands of mass transfer episodes, transferring both mass and angular momentum to the pulsar thus giving a strong hint on the recycling scenario \citep{bhattacharya_1991}. Moreover we obtain higher value for $\dot M_2,\ \dot P_{\rm orb}$ and $\dot \nu_{\rm psr}$ in both point where we observe the sequences, thus, albeit with a grain of salt, suggesting that the source could be in the ascending phase of a cycle. 

In addition to the description of the evolution of \saxj, a number of interesting observations can be made when studying the long term evolution of the models that can be summarized as follows.

The evolution of the models is cyclic along its entire length. This is a known result, outlined for the first time in the seminal paper by \cite{pod_1991} and then explored by many other authors \citep[e.g.:][and references therein]{dantona_1994,buning_2004,benvenuto_2014,benvenuto_2017}.  
When we observe the details of each model, we see that both the shape and the duration of the cycles are different in different segments of the tracks (see Figure \ref{FIG_TVR_RR} and \ref{FIG_LH}), thus we have been able to divide the evolution in three main phases, corresponding to the various shapes and the general behaviour of the tracks in the $M_2$ -- $P_{orb}$ plane.
We identify, on this basis, on the $M_2$ -- $P_{orb}$ plane,  an early phase showing large variation of orbital period and radius of the donor; an intermediate phase where generally, the orbital period tends to increase, while at the same time the cycles are shorter; and a later phase where the short cycles are paired with a general decrease of the orbital period, until it reaches its minimum value. We remark that the cycles we find in the early evolutionary phase are similar to the one found in other works, e.g. \cite{buning_2004,benvenuto_2014,benvenuto_2017}, with the difference that in our model the system becomes completely detached\footnote{Which does not happens in every work mentioned here, see Fig. 24 in \cite{buning_2004}.}
Each cycle includes a short phase where the system transfers mass from the companion star to the neutron star and a longer stage without mass transfer. During the mass transfer phases the irradiation, through the feedback process we described in \S~\ref{EVO_DONOR}, makes the orbital period to increase, while in the detached phase the angular momentum losses by GW and MB shrink down the system. The pulsar spin  (following the evolution choices described in \S 2.2) shows a maximum at $M_2 \sim 0.2\ M_\odot$ (see Figure \ref{FIG_MVNS}) which, in the three models presented, is $550 \leq \nu_{max} \leq 650$. We interpret this as a combination of the increasing efficiency of the radio ejection phenomenon and the consequent lengthening of the cycles, including their detached phase.

The models suggests that the status of the system, i.e. whether it is observed as a LMXB or as a radio pulsar (and, later on, as a radio MSP) is not bound to a specific evolutionary stage. Anyway, looking at the general evolution, as the length of the cycles changes along the tracks (see e.g. Figure \ref{FIG_MDOT}), the probability to find the system in each of the two stages is different, and the pulsar status is the most probable one. 
Moreover, the enhancement of the mass loss due to irradiation has an interesting effect on the evolution: the duration of the three phases differs greatly. Indeed, looking at the numbers reported in Table \ref{TAB_ratio} we see that each of the models spend most of their lifetime (about 75\%) at $M_2/M_\odot \leq 0.2$, and thus we may explain why we see  less system at higher values of $M_2$.

Both sources of irradiation are needed to produce models that offer a good description of the observed orbital period and mass of the system. We showed in Figure \ref{FIG_MVSP} that models calculated including a single contributing source of irradiation (either from X-rays,  or from the  MSP),  do not produce a satisfying result, because the evolution reaches orbital periods too short. In each case, however, we have left out a significant part of the physics involved in these systems. Taking into account both the contributions of the X-ray and MSP irradiation in calculating $L_h$, is an approach usually not followed in the literature \citep[e.g.][]{buning_2004} as the irradiation contribution from the MSP spin down radiation is order of magnitudes below the X-ray irradiation (see Figure \ref{FIG_LH}); nevertheless, as the system, at low donor masses, spends most of its life time detached (see e.g. Figure \ref{FIG_LH}), this contribution becomes crucial as  driving mechanism guiding the orbital evolution of these models. The pulsar irradiation significantly alters the mass - radius relation of the donor (Figure \ref{FIG_MVR}), so that the system stays at larger orbital periods. This effect is directly tied to the relevance of the pulsar luminosity contribution to the irradiation field and is bigger in the models with higher efficiency values (top panel of Figure \ref{FIG_MVSP}). We plan to discuss these points more in detail in a future paper (Tailo et al. in preparation).

The models, because of the effects irradiation has on the donor internal structure, show that the companion star is not fully convective; instead, as we have shown in Figure \ref{FIG_MVMBCE}, even at values of $M_2 \sim 0.065\ M_\odot$ we obtained models showing a non negligible radiative layer. This suggest that the MB braking still plays a role in shaping the evolution of these stars and systems and may be the reason why we do not see the typical period gap among the system containing a neutron star. Our models reflect this finding by keeping MB active along the entire evolution, as in the recent work by \cite{chen_2017}, at variance with the standard model for a CV-like evolution that resorts to pure GW AM losses when the companion star becomes fully convective. 

\section*{Acknowledgements}

We acknowledge financial contribution from the agreement ASI-INAF I/037/12/0 and ASI-INAF 2017-14-H.O. We also acknowledge support from the HERMES Project, financed by the Italian
Space Agency (ASI) Agreement n. 2016/13 U.O, as well as fruitful discussion
with the international team on The disk-magnetosphere interaction around
transitional millisecond pulsars at the International Space Science Institute, Bern.

M.T. acknowledges support from the University of Cagliari and partial support by the European Research Council through the ERC-StG 2016, project 716082 `GALFOR', and by the MIUR through the FARE project R164RM93XW `SEMPLICE'. 
A. P. acknowledges funding from the EUs Horizon 2020 Framework Programme for Research and Innovation under
the Marie Skodowska-Curie Individual Fellowship grant agreement 660657-TMSP-H2020-MSCA-IF-2014




\bibliographystyle{mnras}
\bibliography{1808} 








\bsp    
\label{lastpage}
\end{document}